\documentclass[a4paper]{jpconf}
\usepackage{graphicx}
\begin{document}
\title{Radial velocities of giant stars: an investigation of line profile variations.}

\author{S Hekker$^1$, I A G Snellen$^1$, C Aerts$^{2,3}$, A Quirrenbach$^4$, S Reffert$^4$ and D~S~Mitchell$^5$}

\address{$^1$ Leiden Observatory, Leiden University, P.O. Box 9513, 2300 RA Leiden, The Netherlands}
\address{$^2$ Instituut voor Sterrenkunde, KU Leuven, Celestijnenlaan 200D, 3001 Leuven, Belgium}
\address{$^3$ Department of Astrophysics, University of Nijmegen, P.O. Box 9010, 6500 GL Nijmegen, The Netherlands}
\address{$^4$ ZAH, Landessternwarte Heidelberg, K\"onigstuhl 12, D-69117 Heidelberg, Germany}
\address{$^5$ California Polytechnic State University, San Luis Obispo, CA 93407, USA}

\ead{saskia@strw.leidenuniv.nl}

\begin{abstract}
Since 1999, a radial velocity survey of 179 red giant stars is ongoing at Lick Observatory with a one month cadence.  At present $\sim$20$-$100 measurements have been collected per star with an accuracy of 5 to 8 m\,s$^{-1}$. Of the stars monitored, 145 (80\%) show radial velocity (RV) variations at a level $>$20 m\,s$^{-1}$, of which 43 exhibit significant periodicities. Here, we investigate the mechanism causing the observed radial velocity variations. Firstly, we search for a correlation between the radial velocity amplitude and an intrinsic parameter of the star, in this case surface gravity ($\log g$). Secondly, we investigate line profile variations and compare these with theoretical predictions.
\end{abstract}

\section{Introduction}
Since 1999, a radial velocity survey of 179 red giant stars is ongoing at Lick Observatory, using the 60 cm Coud\'e Auxiliary Telescope (CAT) in conjunction with the Hamilton echelle spectrograph (R $\approx$ 60\,000).  These stars have been selected from the Hipparcos catalogue \cite{esa1997}, based on the criteria described by \cite{frink2001}. The selected stars are all brighter than 6~mag, are presumably single and have photometric variations $< 0.06$~mag in V.  The system with an iodine cell in the light path has been developed as described by \cite{marcy1992} and \cite{valenti1995}. With integration times of up to thirty minutes for the faintest stars ($m_{v}$ = 6 mag) we reach a signal to noise ratio of about $80-100$ at $\lambda = 5500$ \AA , yielding a radial velocity precision of $5-8$ m\,s$^{-1}$. 

The initial aim of the survey was to check whether red giants would be stable enough to serve as reference stars for astrometric observations with SIM/PlanetQuest \cite{frink2001}. In \cite{hekker2006} it is shown that a large fraction of the red giants in a specific part of the absolute magnitude vs. B-V colour diagram are stable to a level of 20 m\,s$^{-1}$ and could be effectively searched for long period companions, as is required for astrometric reference stars. For other stars in the sample the radial velocity variations are larger, and for 43 stars these show significant periodicities. So far, sub-stellar companions have been announced for two stars from the present sample ($\iota$ Dra \cite{frink2002} and $\beta$ Gem \cite{reffert2006}). 

Here, we investigate which physical mechanism causes the observed radial velocity variations. In cases for which we do not find a significant periodicity in the observed radial velocity variations, an intrinsic mechanism such as spots or pulsations, possibly multi-periodic, seems most likely. On the other hand, the periodic radial velocity variations can be caused by sub-stellar companion, an intrinsic mechanism, or by both these mechanisms simultaneously. In Section 2 we search for a relation between the amplitude of the radial velocity variations and an intrinsic parameter, i.e.~$\log g$. In Section 3 we investigate line shape variations and compare these with theoretical predictions. Our conclusions are presented in Section 4. A more extended paper on this subject is submitted \cite{hekker2007b}.

\section{Radial velocity amplitude vs.~surface gravity relation}
\begin{figure}
\begin{center}
\includegraphics[width=\linewidth]{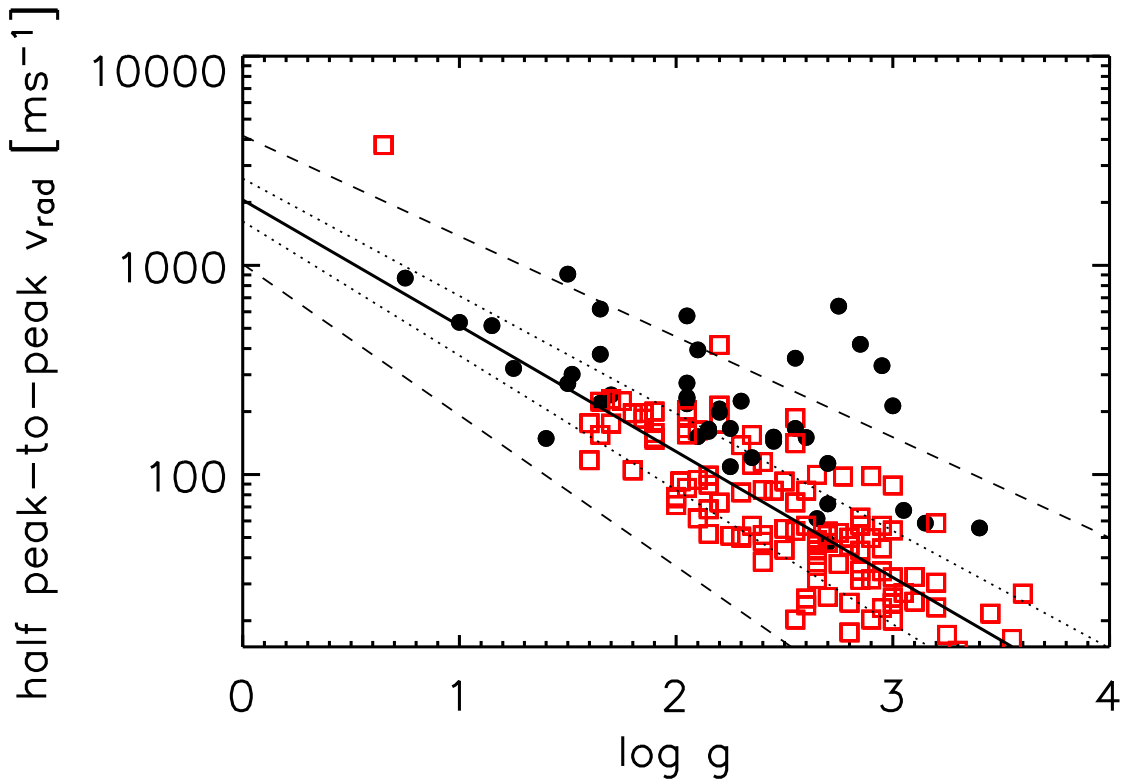}
\caption{\label{klogg}Half of the peak-to-peak variation of the radial velocity as a function of surface gravity ($\log g$). The black dots indicate the stars with periodic radial velocity variations (stellar binaries are excluded), and the red squares indicate stars with random radial velocity variations. The solid line is the best fit through the random stars, the dotted line indicates the $1\sigma$ interval around the best fit and the dashed line indicates the $3\sigma$ interval. Six of the 8 stars with periodic radial velocity variations and $\log g < 1.6$ are classified bright giants or supergiants \cite{esa1997}.}
\end{center}
\end{figure}

Hatzes and Cochran 1998 \cite{hatzes1998} already investigated the origin of the observed radial velocities in K giant stars. Although their sample contained only 9 stars, they suggested that the amplitude of the radial velocity increases with decreasing surface gravity.  In lower surface gravity it takes longer to decrease the velocity of a moving parcel which results in larger amplitudes and the relation suggested by \cite{hatzes1998} would therefore be evidence for an intrinsic mechanism for these long period radial velocity variations.

For the present sample, $\log g$ values were determined spectroscopically by \cite{hekker2007}, by imposing excitation and ionisation equilibrium of iron lines through stellar models. The equivalent width of about
two dozen carefully selected iron lines were used for a spectroscopic LTE analysis based on the 2002 version of MOOG \cite{sneden1973} and Kurucz model atmospheres which include overshoot effects \cite{castelli1997}. These authors estimated the error on $\log g$ to be 0.22 dex from the scatter found in a
comparison with literature values. A detailed description of the stellar parameters for individual stars and a comparison with literature values are available in \cite{hekker2007} and is therefore omitted here.

\begin{figure}
\begin{center}
\includegraphics[width=\linewidth]{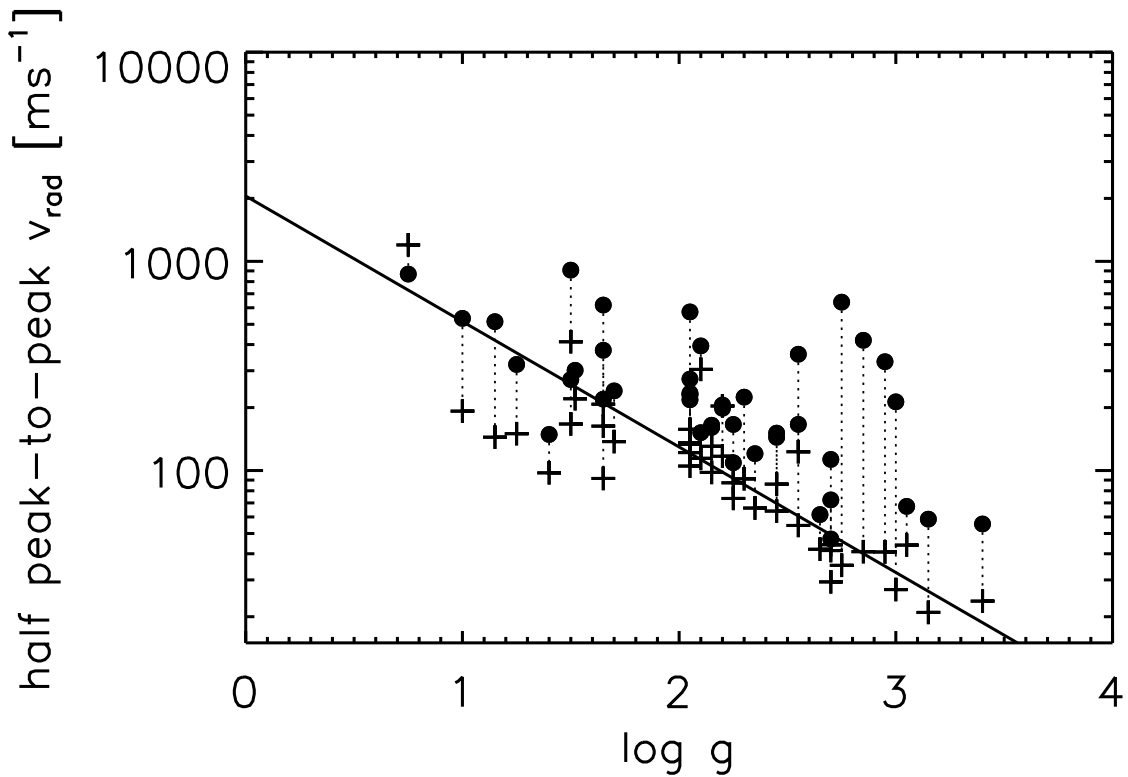}
\caption{\label{kloggres}Half of the peak-to-peak variation of the radial velocity as a function of surface gravity ($\log g$) as in Figure~\ref{klogg}, but showing only those stars with periodic radial velocity variations (dots) and their residual (plus). The solid line indicates the linear fit through the stars with non-periodic radial velocity variations (from Figure~\ref{klogg}).}
\end{center}
\end{figure}

In Figure~\ref{klogg} the half peak-to-peak value of the radial velocity is plotted as a function of $\log g$, together with the best linear fit. There clearly exists a correlation between the observed radial velocity amplitude variation and the surface gravity. Also, most of the stars with periodic radial velocity variations and $\log g > 1.6$ dex are located above the best fit, which could indicate that both intrinsic and extrinsic mechanisms are contributing. To investigate this, we subtracted the periodic fit from the radial velocity variations and plotted half the peak-to-peak value of the residuals as a function of surface gravity, see Figure~\ref{kloggres}. For stars with $\log g > 1.6$ dex the residuals are now located around the best fit through the random stars. This is an indication that intrinsic and extrinsic mechanisms are indeed contributing simultaneously in these stars. 

For 8 out of 9 stars with $\log g < 1.6$ dex, we find a significant period and therefore, the best fit in Figure~\ref{klogg} may not be very accurate in this region. Furthermore, the atmospheres of stars with these low surface gravities are so diluted that instabilities occur easily, either periodic or random. We therefore think it most likely that these variations are not due to companions. Moreover, the radial velocity variations of these stars are located around the fit for random variables, while the residuals are mostly below this relation.

\section{Line shape analysis}
In the previous section we treated a sample of stars, but we would also like to know what mechanism causes the radial velocity variations  in each star individually. For a sub-sample of stars, we therefore obtained high resolution spectra (R $\approx$ 164\,000) with the SARG spectrograph mounted on the Telescopio Nazionale Galileo, La Palma, Spain. We have between 3 and 8 observations per star, which is not enough to do a full line shape analysis. But we tried to identify whether significant line depth variation is present in a star, which would indicate an intrinsic mechanism. To do this we shifted the spectra of each star, taken at different epochs, to the laboratorium wavelength and computed a time averaged profile. Residuals at each epoch provides us with the variation in line depth which is  indicative of line shape variations and thus the presence of an intrinsic mechanism.

\begin{figure}
\begin{center}
\begin{minipage}{18pc}
\includegraphics[width=18pc]{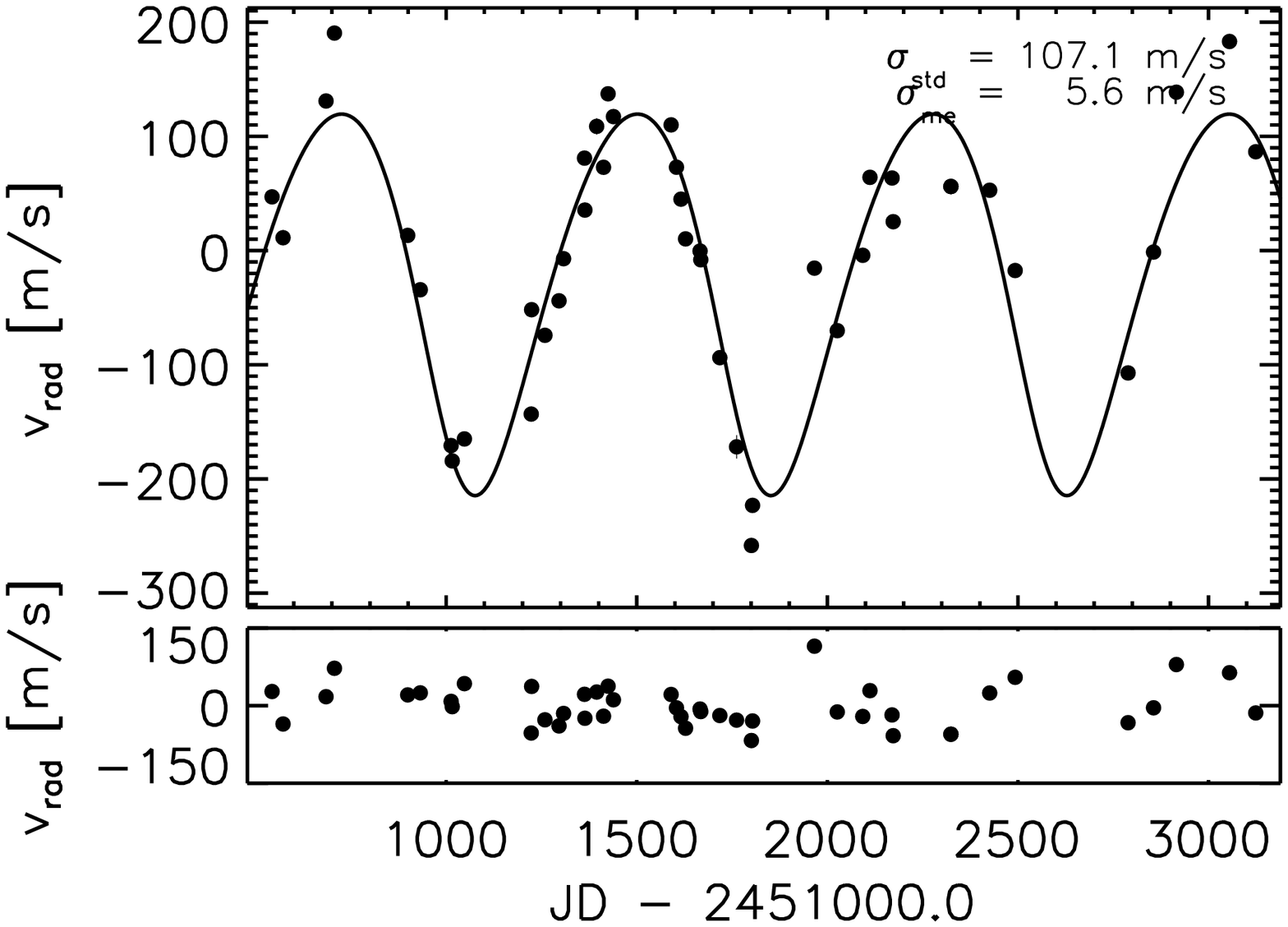}
\end{minipage}\hfill
\begin{minipage}{18pc}
\includegraphics[width=18pc]{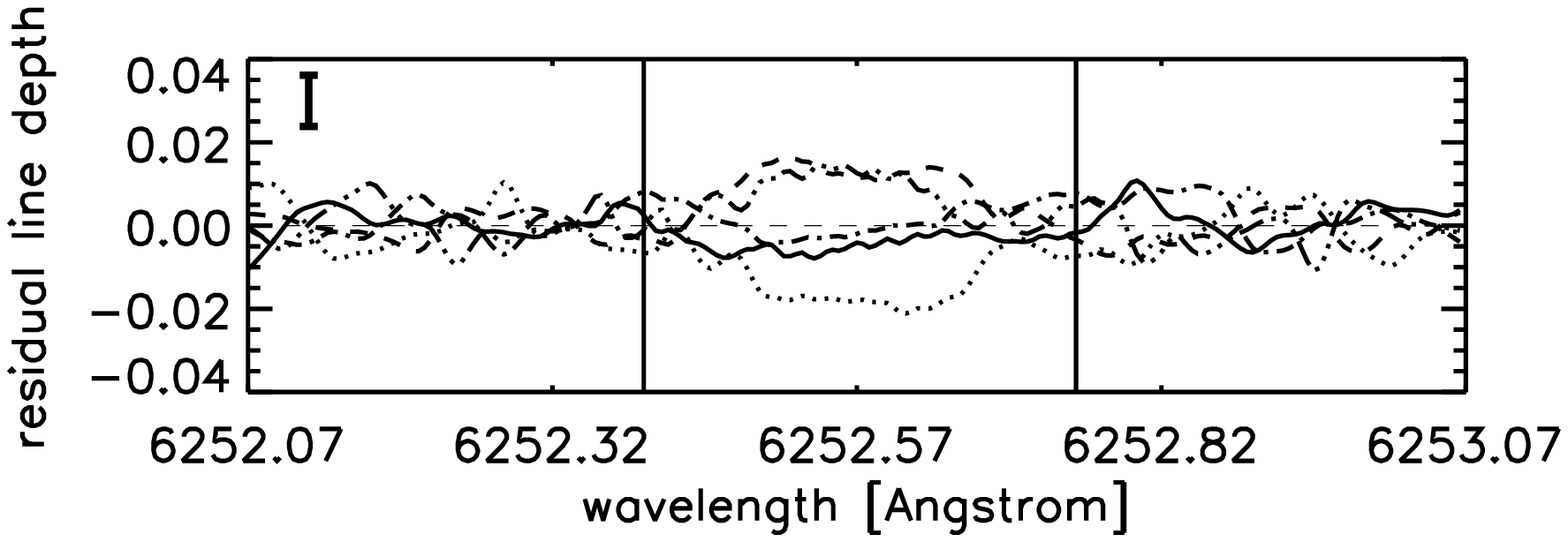}
\end{minipage}
\caption{\label{hip53261} Left: radial velocity variations of HIP53261 as a function of Julian Date. A Keplerian orbit is fitted through the data and the residuals are shown in the bottom panel. Right: the residual of the Fe I line at 6252.57 \AA , taken at different epochs (indicated with different line styles), with respect to a time averaged profile, as a function of wavelength. The vertical lines in this panel indicate the spectral line in wavelength. An error estimate is indicated with the thick error bar in the left upper corner. }
\end{center}
\end{figure}

In Figure~\ref{hip53261} we show the radial velocity variation of HIP53261 as a function of phase and the residuals of the line depth. For this star we see significant variation in the line depth, which is indicative of an intrinsic mechanism. Because we lack data to perform a frequency analysis it is not yet possible to verify whether the period of the radial velocity and line depth variation are related, in which case the variations are due to an intrinsic mechanism. In case the periods are not related, we probably have a companion orbiting an intrinsically active star.

\begin{figure}
\begin{center}
\begin{minipage}{18pc}
\includegraphics[width=18pc]{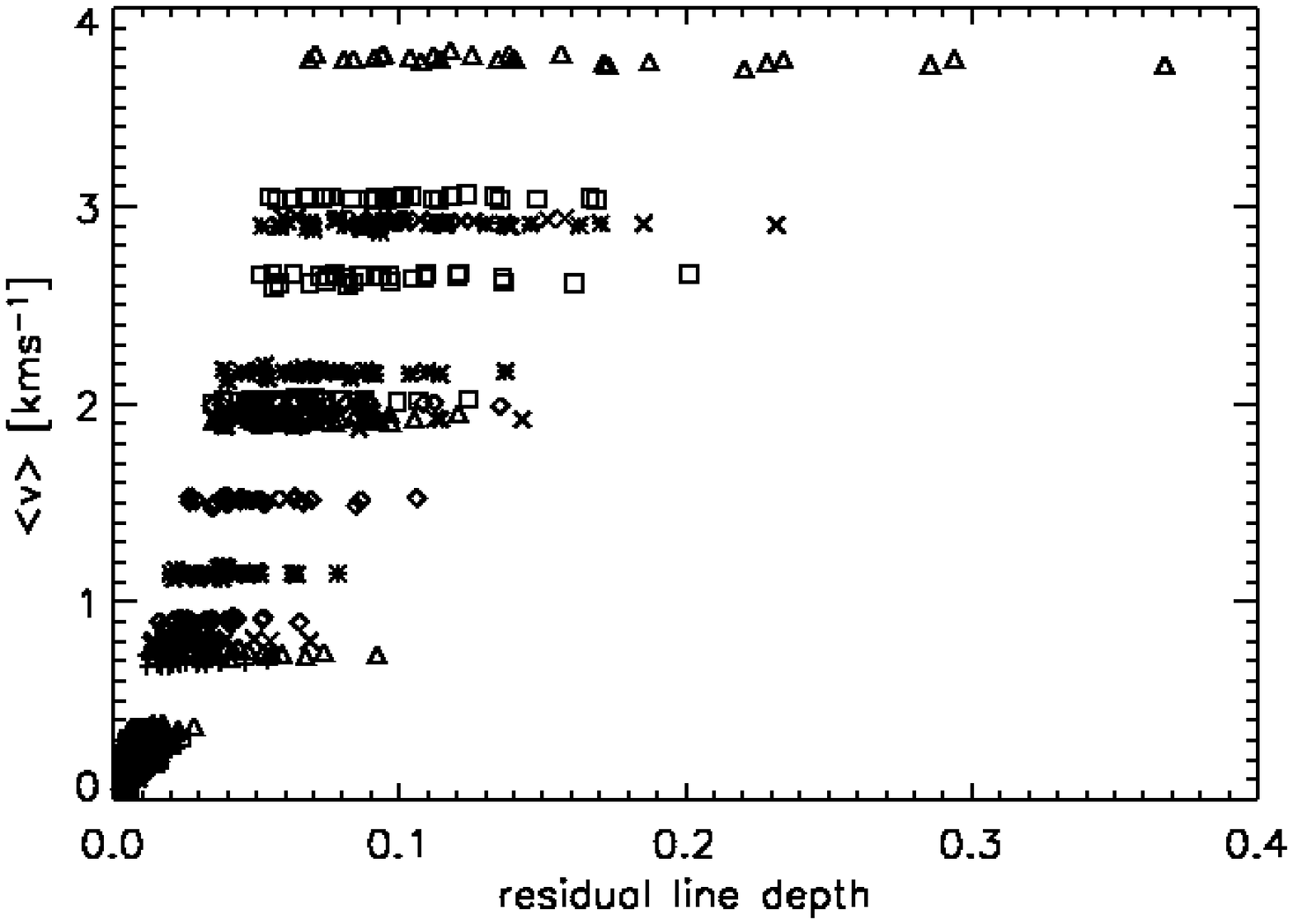}
\end{minipage}\hfill
\begin{minipage}{18pc}
\includegraphics[width=18pc]{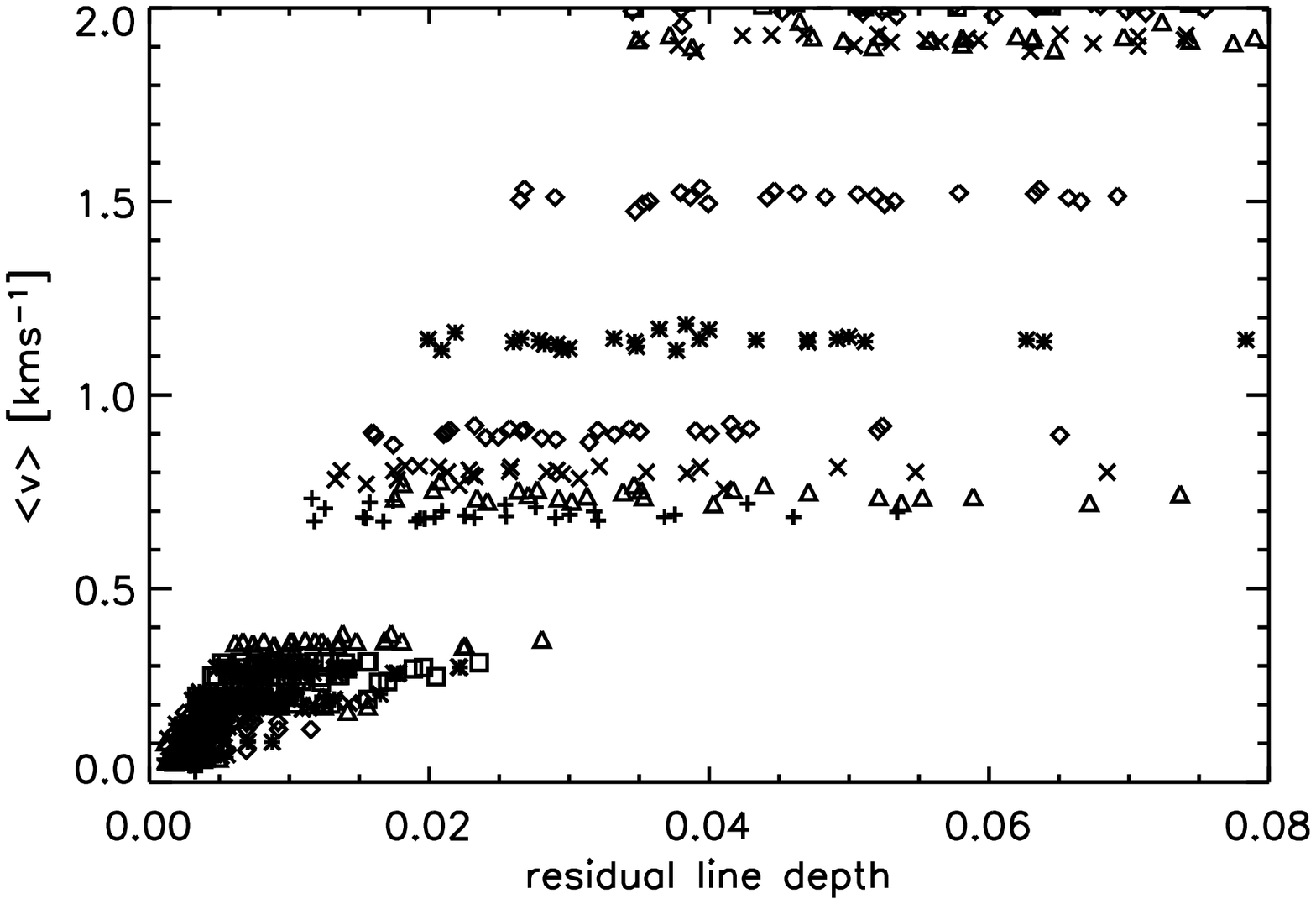}
\end{minipage}\hfill
\begin{minipage}{18pc}
\includegraphics[width=18pc]{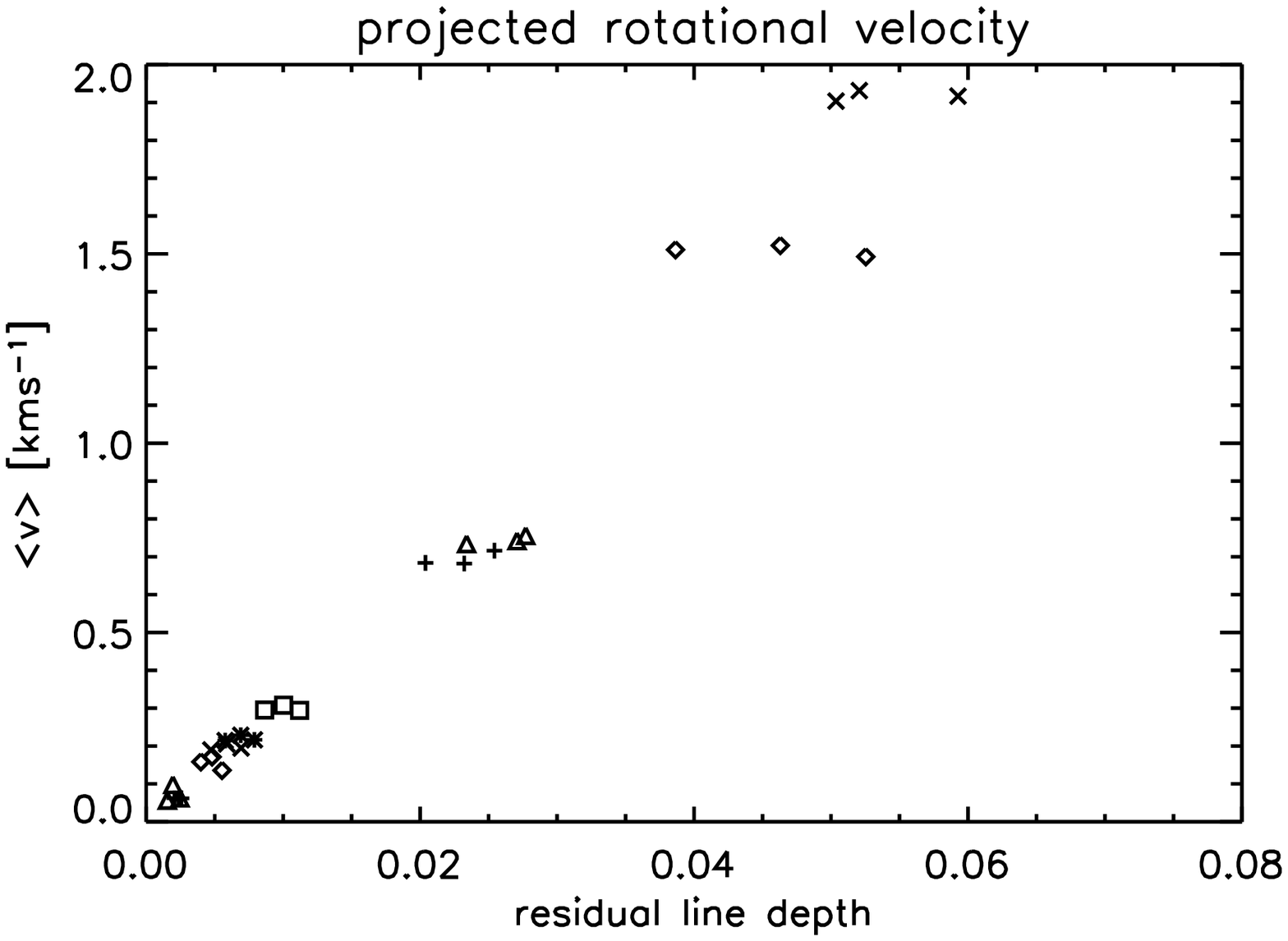}
\end{minipage}\hfill
\begin{minipage}{18pc}
\includegraphics[width=18pc]{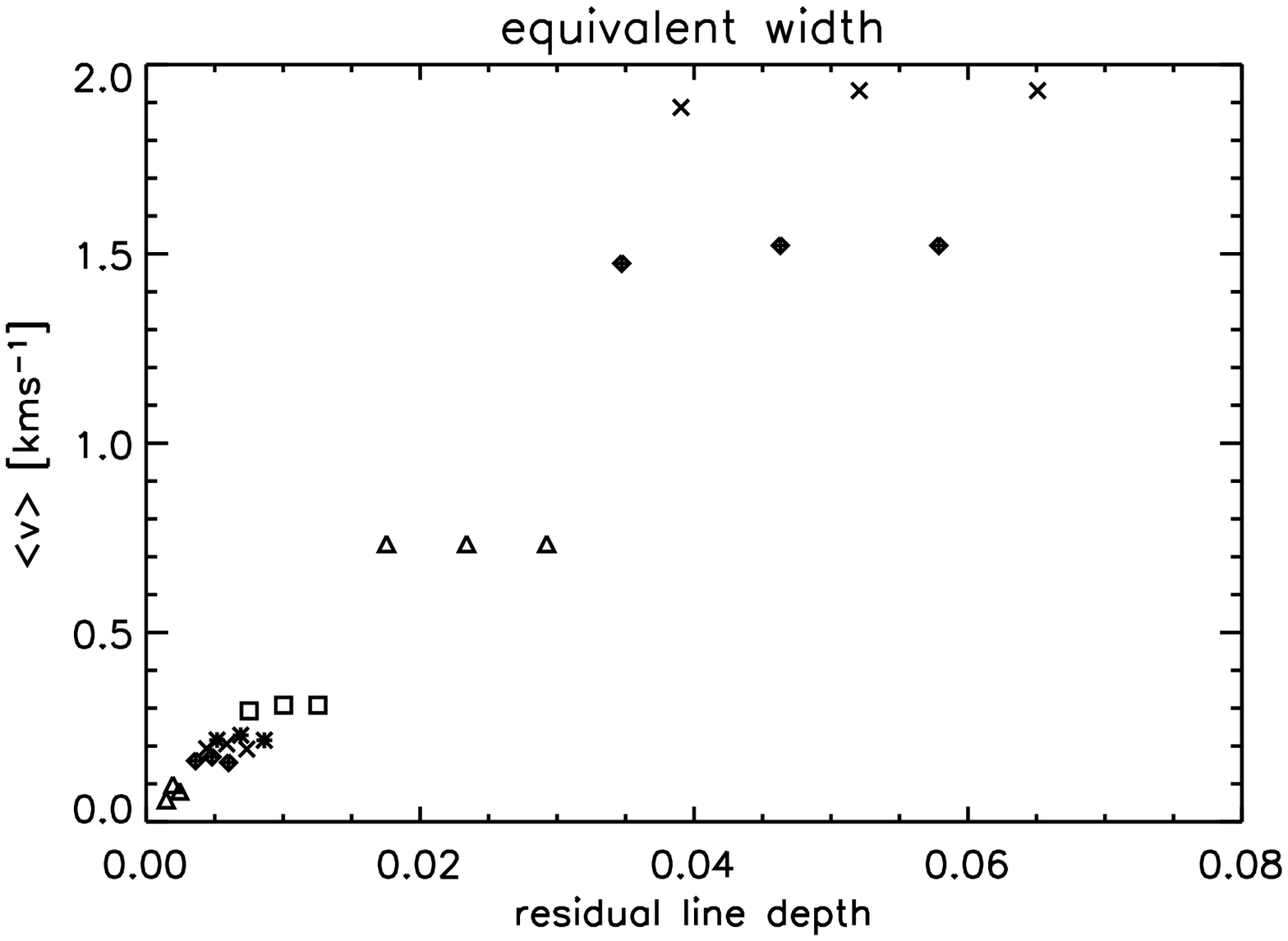}
\end{minipage}\hfill
\begin{minipage}{18pc}
\includegraphics[width=18pc]{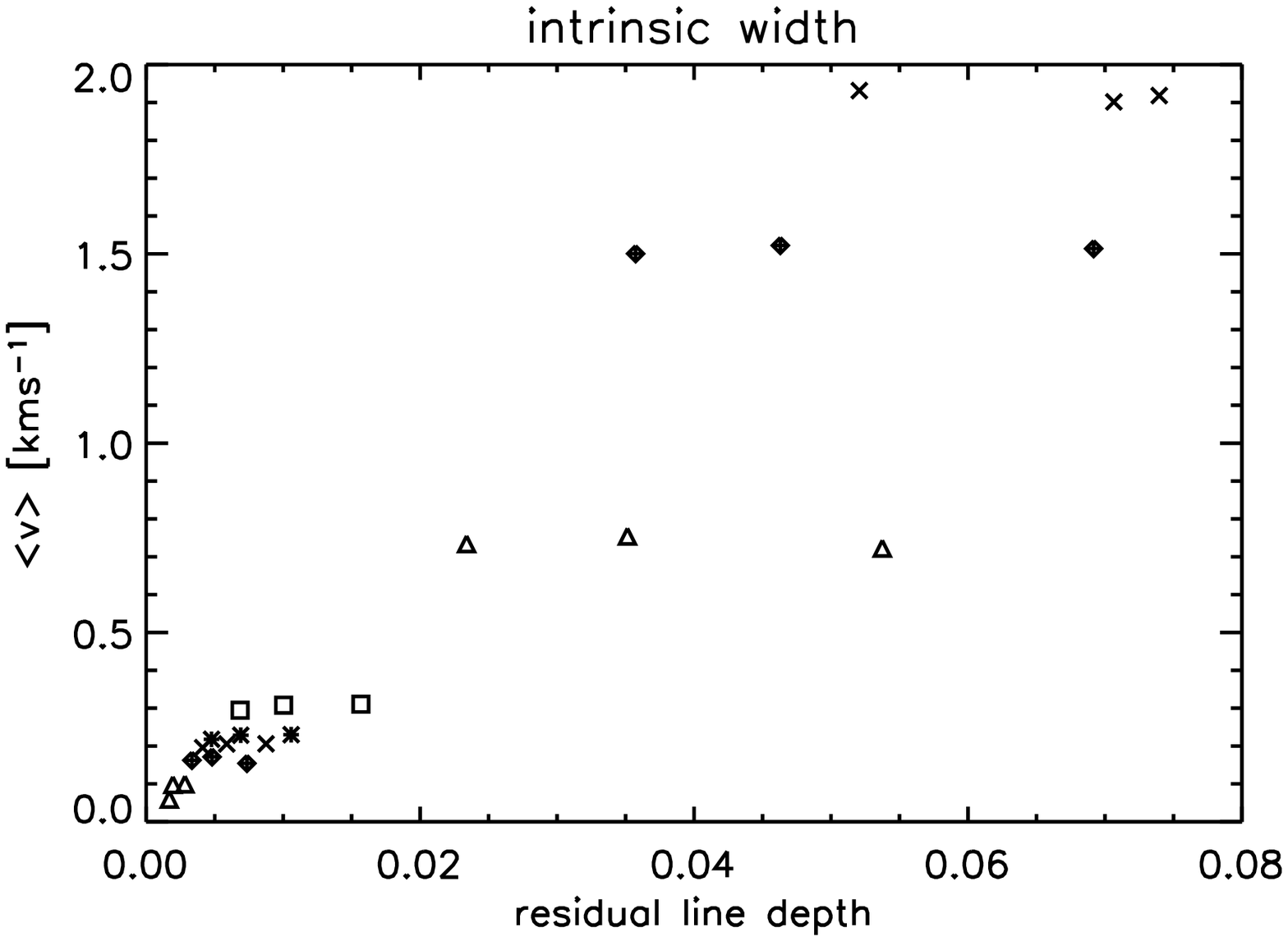}
\end{minipage}\hfill
\begin{minipage}{18pc}
\includegraphics[width=18pc]{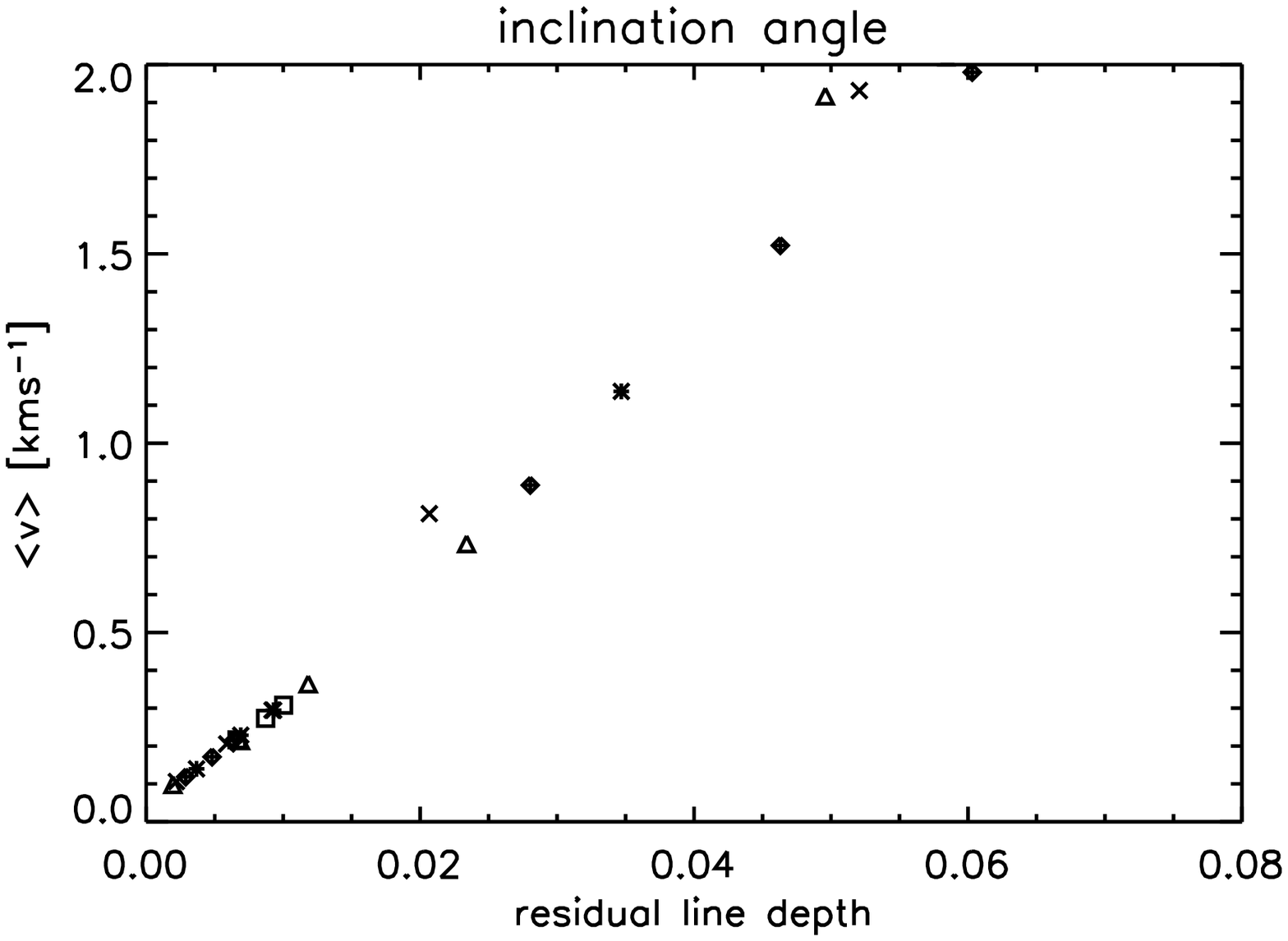}
\end{minipage}\hfill
\begin{minipage}{18pc}
\includegraphics[width=18pc]{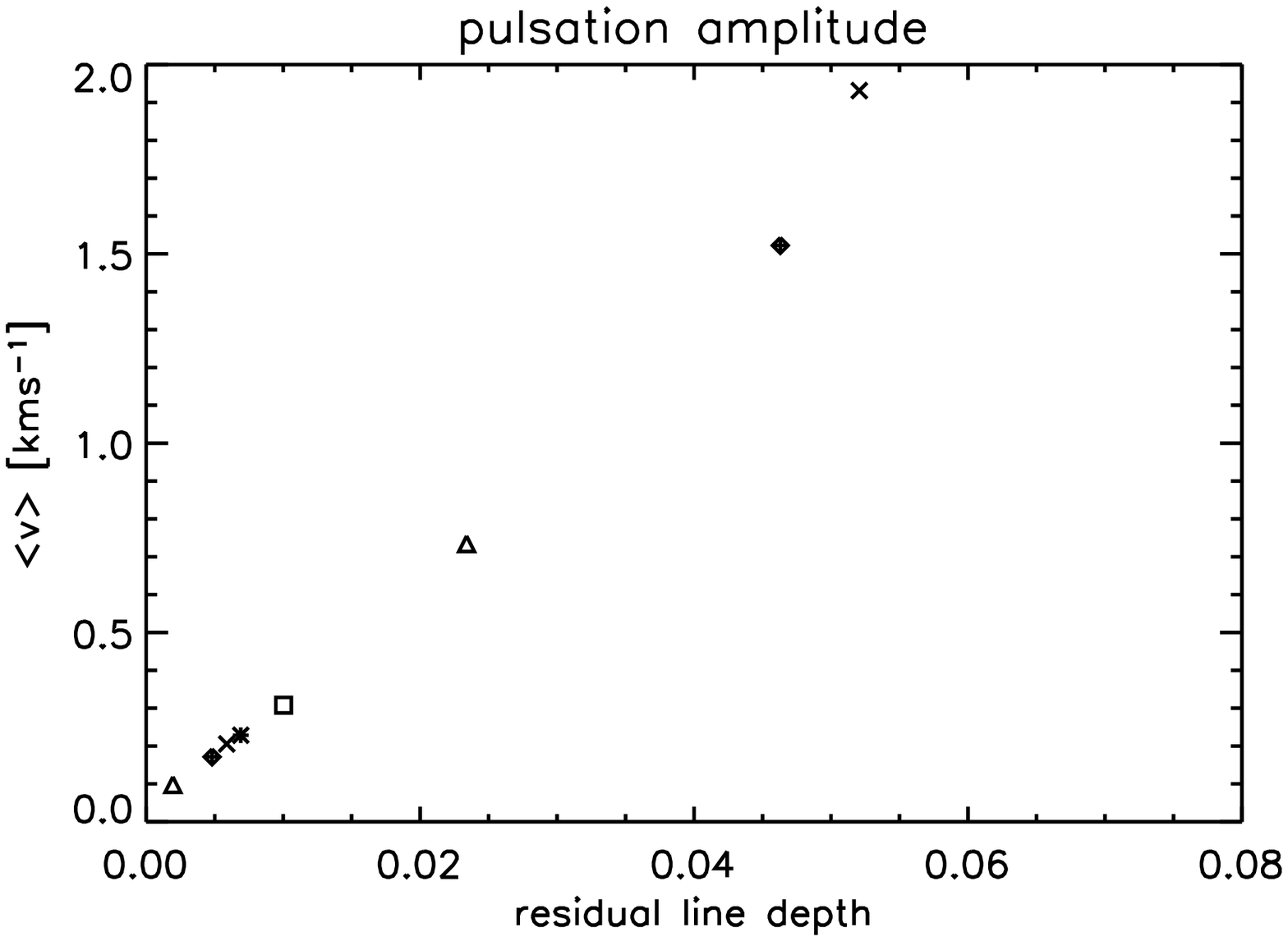}
\end{minipage}\hfill
\begin{minipage}{18pc}
\caption{\label{ressim} Half peak-to-peak values of the first moment as a function of residuals in line depth for stars with pulsations with $\ell=(0,1,2)$, positive $m$ values (different modes are indicated with different symbols). Top: for inclination angles of 30, 50 and 70 degrees, projected rotational velocity of 2.0, 3.5 and 5.0 km\,s$^{-1}$, equivalent width of 30, 40 and 50 km\,s$^{-1}$, intrinsic  line width of 3.0, 4.0 and 5.0 km\,s$^{-1}$ and pulsation amplitudes of 0.1 and 1.0 km\,s$^{-1}$. Top: all computed points. Others from left to right and top to bottom: dependence on projected rotational velocity, equivalent width, intrinsic line width, inclination angle and pulsation amplitude.}
\end{minipage}
\end{center}
\end{figure}

\begin{figure}
\begin{center}
\includegraphics[width=\linewidth]{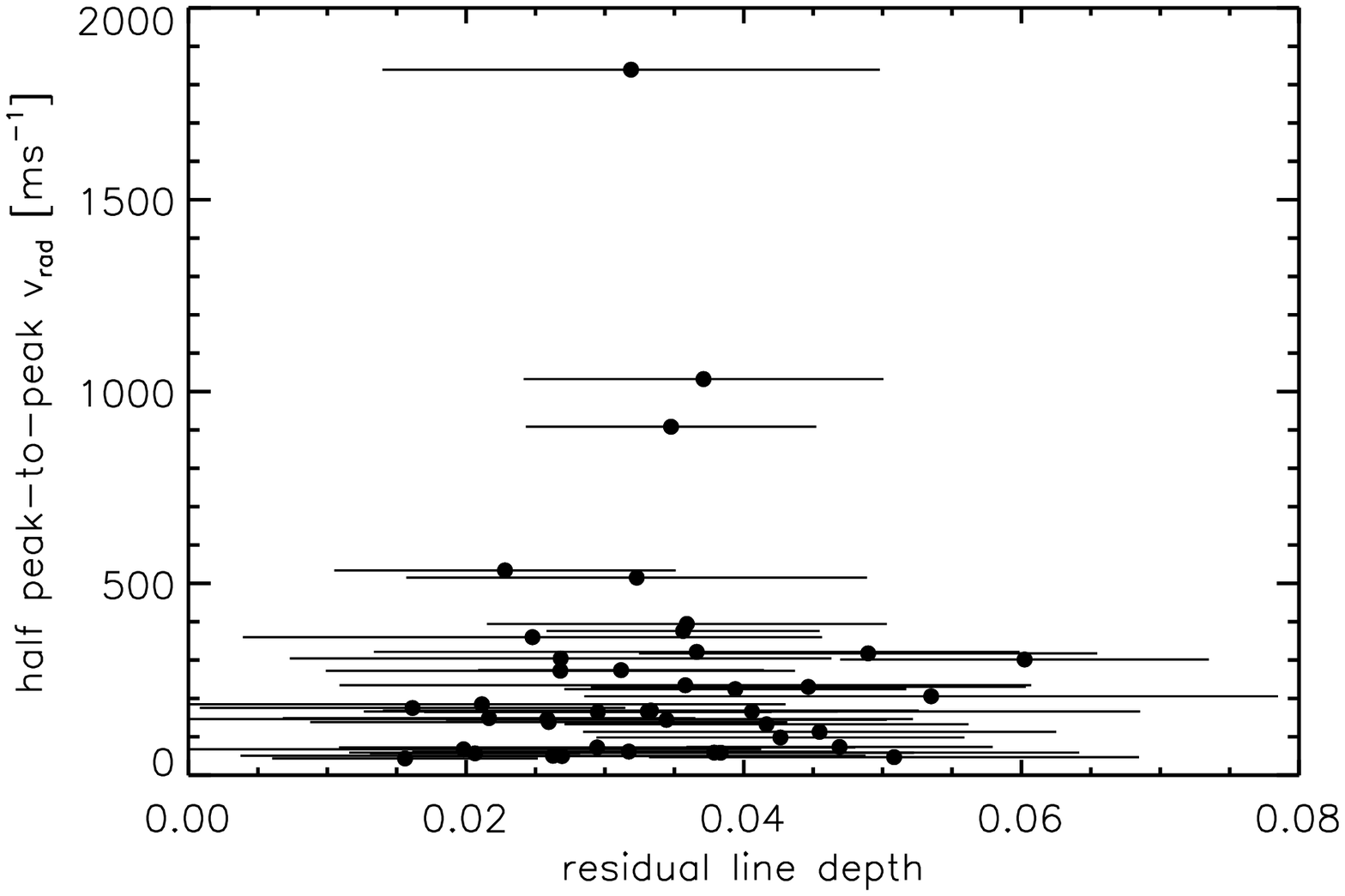}
\caption{\label{RVdif}Half of the peak-to-peak variation of the radial velocity as a function of residuals in line depth, for all stars with at least 3 high resolution, high signal to noise observations with SARG.}
\end{center}
\end{figure}

Apart from the relation in frequency between the radial velocity and line depth variations we are also interested in the possible existence of a correlation between the amplitudes of both variations. Because the stars show hardly any photometric variation, spots similar to sunspots are not the most likely mechanism and therefore we focus here on stellar pulsations. We computed the half peak-to-peak value of the first moment (another diagnostic for the radial velocity) and residuals in line depth for line profiles with pulsation modes $\ell=0, 1, 2$ and positive $m$ values, inclination angles of 30, 50 and 70 degrees, projected rotational velocities of 2.0, 3.5 and 5.0 km\,s$^{-1}$, intrinsic width of 3.0, 4.0 and 5.0 km\,s$^{-1}$, equivalent width of 30, 40 and 50 km\,s$^{-1}$ and pulsation velocities of 0.1 and 1.0 km\,s$^{-1}$, and plotted these in the top panels of Figure~\ref{ressim}. From these plots it is clear that there is no simple relation between the amplitude in the radial velocity and residual line depth variations. To investigate the influence of different parameters, we also plotted the first moment amplitude as a function of residual line depth for all modes and both pulsation amplitudes, and vary one parameter, while keeping three other parameters fixed (Figure~\ref{ressim}). In this way it becomes clear that the projected rotational velocity, intrinsic line width and equivalent width of a line introduces variations in the line residuals at approximately the same value for the first moment. On the other hand, varying the inclination angle or pulsation velocity changes the amplitude of the first moment (radial velocity variation) at approximately the same residual line width. 

In Figure~\ref{RVdif} we show the observed half peak-to-peak values of the radial velocity variation as a function of the residual line depth variation, for all stars with at least 3 high resolution, high signal to noise observations with SARG. There is no correlation between these points, which is as expected from the theoretical analysis for stars with different parameters. Strikingly, most of the observed points fall in a region where no theoretical values are found. This difference between the observations and computations can be caused by the fact that in the theoretical models the temperature variation of the stellar surface due to oscillations is not taken into account. This temperature variation causes variations in the intrinsic and equivalent width of the spectral line and these parameters have a large influence on the amplitude of the residual line depth. This effect could cause the higher amplitudes in the observed line depth residuals than expected from the calculations. In case the input values chosen for the models are too far off the actual values of the pulsation parameters of the stars, we would also expect a discrepancy between the observed and computed relation between radial velocity variations and residual line depth.

\section{Conclusions}
There exists a clear correlation between the half peak-to-peak values of the radial velocity variations and the surface gravity of the red giant stars. This is a strong indication that the observed radial velocity variations are caused by a mechanism intrinsic to the star.  Companions and an intrinsic mechanism might be present in stars with periodic radial velocity variations and $\log g > 1.6$ dex, while for stars with $\log g<1.6$ dex solely an intrinsic mechanism seems most likely.

We have investigated whether there is a relation between the radial velocity amplitude and the amplitude of the line depth residuals. From theory we find that the rotational velocity, intrinsic and equivalent line width largely influence the line depth residual and no clear correlation could be identified.

In the theoretical models temperature variations due to the pulsations is ignored, but these temperature variations influence the intrinsic and equivalent line width of the spectral line, which influences the line depth residuals. This might be an explanation why the theoretical models do not overlap with the observations.


\section*{References}
\begin{thebibliography}{50}
\bibitem{castelli1997} Castelli F, Gratton R G and Kurucz R L 1997 {\it A\&A} {\bf 526} 432

\bibitem{esa1997} Perryman M A C and ESA 1997 {\it ESA Special Publication} {\bf 1200}

\bibitem{frink2001} Frink S, Quirrenbach A, Fischer D A, R\"oser S and Schildbach E 2001 {\it PASP} {\bf 113} 173-87

\bibitem{frink2002} Frink S, Mitchell D S, Quirrenbach A, Fischer D A, Marcy G W and Butler R P 2002 {\it ApJ} {\bf 576} 478-84

\bibitem{hatzes1998} Hatzes A P and Cochran W D 1998 {\it Astronomical Society of the Pacific Conference Series} {\bf 154} 311

\bibitem{hekker2006} Hekker S, Reffert S, Quirrenbach A, Mitchell D S, Fischer D A, Marcy G W and Butler R P 2006 {\it A\&A} {\bf 454} 943-9

\bibitem{hekker2007} Hekker S and Mel\'endez J 2007 {\it A\&A in press}

\bibitem{hekker2007b} Hekker S, Snellen I A G, Aerts C, Quirrenbach A, Reffert S and Mitchell D S 2007 {\it A\&A submitted}

\bibitem{marcy1992} Marcy G W and Butler R P 1992 {\it PASP} {\bf 104} 270-7

\bibitem{reffert2006} Reffert S, Quirrenbach A, Mitchell D S, Albrecht S, Hekker S, Fischer D A, Marcy G W and Butler R P 2006 {\it Apj} {\bf 652} 661-5

\bibitem{sneden1973} Sneden C A 1973 {\it PhD thesis}

\bibitem{valenti1995} Valenti J A, Butler R P and Marcy G W 1995 {\it PASP} {\bf 107} 966
\end{thebibliography}
\end{document}